\documentclass[10pt,conference]{IEEEtran}
\IEEEoverridecommandlockouts

\usepackage{cite}
\usepackage{amsmath,amssymb,amsfonts}
\usepackage{algorithmic}
\usepackage{algorithm}
\usepackage{graphicx}
\usepackage{textcomp}
\usepackage{xcolor}
\usepackage{xcolor, colortbl}
\usepackage[table]{xcolor}
\usepackage[dvipsnames]{xcolor}
\usepackage{multirow}
\usepackage{url}
\renewcommand{\thetable}{\arabic{table}}
\usepackage{caption}
\renewcommand{\figurename}{Figure}
\renewcommand{\tablename}{Table} 

\def\BibTeX{{\rm B\kern-.05em{\sc i\kern-.025em b}\kern-.08em
    T\kern-.1667em\lower.7ex\hbox{E}\kern-.125emX}}
\usepackage{booktabs}

\newcommand{\werSub}[2]{\text{#1}$_{\text{\tiny #2}}$}

\begin{document}

\title{ZO-ASR: Zeroth-Order Fine-Tuning of Speech Foundation Models without Back-Propagation\\
}


\author{
\IEEEauthorblockN{
    Yuezhang Peng\IEEEauthorrefmark{2}, 
    Yuxin Liu\IEEEauthorrefmark{2},
    Yao Li\IEEEauthorrefmark{5}, 
    Sheng Wang\IEEEauthorrefmark{2}\IEEEauthorrefmark{5},
    Fei Wen\IEEEauthorrefmark{3}\IEEEauthorrefmark{1}, 
    Xie Chen\IEEEauthorrefmark{2}\IEEEauthorrefmark{6}\IEEEauthorrefmark{4}\IEEEauthorrefmark{1} 
    }

\IEEEauthorblockA{\IEEEauthorrefmark{2}\textit{X-LANCE Lab, School of Computer Science, MoE Key Lab of Artificial Intelligence, Shanghai Jiao Tong University}}
\IEEEauthorblockA{\IEEEauthorrefmark{3}\textit{School of Information Science and Electronic Engineering/School of Integrated Circuits, Shanghai Jiao Tong University}}
 \IEEEauthorblockA{\IEEEauthorrefmark{6}\textit{Shanghai Innovation Institute} \IEEEauthorrefmark{4}\textit{Jiangsu Key Lab of Language Computing} \IEEEauthorrefmark{5}\textit{Shanghai Aviation Electric Co., Ltd}\\
 \IEEEauthorblockA{
 \{pengyuezhang, chenxie95\}@sjtu.edu.cn}
} 

\thanks{
  Corresponding authors are Xie Chen\IEEEauthorrefmark{1} and Fei Wen\IEEEauthorrefmark{1}.
}
}


\maketitle

\begin{abstract}
Fine-tuning pre-trained speech foundation models for Automatic Speech Recognition (ASR) is prevalent, yet constrained by substantial GPU memory requirements.  We introduce ZO-ASR, a memory-efficient Zeroth-Order (ZO) method that avoids Back-Propagation (BP) and activation memory by estimating gradients via forward passes. When combined with SGD optimizer, ZO-ASR-SGD fine-tunes ASR models using only inference memory. Our evaluation spans supervised and unsupervised tasks. For Supervised Domain Adaptation on Whisper-Large-V3, ZO-ASR's multiple query mechanism enhances robustness and achieves up to an 18.9\% relative Word Error Rate reduction over zero-shot baselines, outperforming existing ZO methods. For unsupervised Test-Time Adaptation on Wav2Vec2-Base, ZO-ASR exhibits moderately lower performance compared to first-order optimizer Adam. Our BP-free approach provides a viable solution for fine-tuning ASR models in computationally resource-constrained or gradient-inaccessible scenarios. \footnote{Our code is available at \url{https://github.com/Gatsby-web/ZO-ASR}.}
\end{abstract}


\section{Introduction}

Influenced by scaling laws in deep learning \cite{kaplan2020scaling}, the parameter numbers and training data volumes of ASR foundation models have continuously increased in recent years. This evolution is evident in the progression from the Wav2Vec2 series (317M parameters trained on 53k hours data) \cite{baevski2020wav2vec} and the HuBERT series (964M parameters trained on 60k hours data) \cite{hsu2021hubert} to the Whisper series (1550M parameters trained on 680k hours data) \cite{radford2023robust}. Currently, the state-of-the-art ASR foundation models, such as Whisper-Large-v3, have surpassed 1.5 billion parameters, achieving enhanced robustness, generalization capabilities, and multilingual recognition abilities.

As model sizes grow, the substantial memory required for fine-tuning large models has become a primary challenge for downstream task adaptation. For instance, fine-tuning 1.55B parameters of Whisper-Large-V3 using the Adam optimizer necessitates at least 40GB GPU memory \cite{adam}, which can easily exceed the memory limits of one consumer-grade GPU.

The common approach to address memory constraints is Parameter-Efficient Fine-Tuning (PEFT) \cite{lester2021power}, encompassing techniques like Prefix-Tuning \cite{li2021prefix} and Low-Rank Adaptation (LoRA) \cite{hu2021lora}. While these methods fine-tune a few model parameters using First-Order (FO) gradient optimizers (e.g., SGD \cite{amari1993backpropagation} and Adam) with back-propagation, the GPU memory required can still be several times that needed for inference. This is because activation values must be stored during gradient computation in back-propagation.

In this paper, we propose ZO-ASR, introducing zeroth-order (ZO) optimization methods to ASR for the first time. Our approach is based on the Memory-Efficient Zeroth-Order Optimization (MeZO) method \cite{malladi2023fine}, which estimates gradients through two forward passes, thereby avoiding back-propagation and eliminating the need to store activation values. ZO-ASR introduces an additional number of queries, reducing the training instability and poor robustness that MeZO faces when fine-tuning ASR models. When combined with the SGD optimizer, ZO-ASR-SGD enables LLM fine-tuning using only inference memory. We further extend this method to the Test-Time Adaptation (TTA) \cite{lin2022listen}, where ZO-ASR method allows the model to adapt to noisy or out-of-domain datasets through only forward passes. This offers a viable solution for adapting ASR models in edge deployment or post-quantization scenarios where back-propagation is infeasible. In summary, our main contributions are as follows:

\begin{itemize}
    \item We propose the ZO-ASR method. Compared to existing MeZO algorithms, mainly focused on fine-tuning Large Language Models (LLMs) on simple classification tasks, ZO-ASR applies additional queries (q-RGE method) to reduce gradient estimation noise, thereby enhancing the robustness of the optimization algorithm.
    \item We conducted experiments on Whisper models of various sizes and demonstrated that ZO-ASR achieves superior performance over both zero-shot and other ZO baselines, while maintaining significantly lower memory requirements than FO optimizers.
    \item We further introduce the ZO-ASR method into the TTA domain, achieving performance with a moderate degradation compared to FO optimizers in experiments involving noisy and out-of-domain datasets. Furthermore, ZO-ASR allows models to adapt solely through forward passes, offering a novel solution for TTA in scenarios where back-propagation is not possible due to edge deployment or model quantization.
\end{itemize}

\section{Related Work}

\subsection{ASR Foundation Model Fine-Tuning}

Pre-training speech foundation models on large-scale datasets and then fine-tuning them for downstream tasks has become the dominant paradigm in speech recognition, achieving state-of-the-art performance across various domains \cite{wu2023decoder,song2024lora, wang2023whislu, child}. For example, fine-tuning Whisper with extensive child speech data significantly improved ASR accuracy for children \cite{child}, while adapting it with spoken language understanding data enabled intent understanding performance surpassing baseline systems \cite{wang2023whislu}. Another common approach is to use LoRA for parameter- and memory-efficient fine-tuning. LoRA-Whisper \cite{song2024lora} fine-tunes the LoRA modules using the MLS \cite{pratap2020mls} and FLEURS \cite{conneau2023fleurs} multilingual datasets, achieving relative gains of 18.5\% in multilingual ASR. However, fine-tuning the 1.55B parameters Whisper-Large-V3 model with LoRA still demands over 30GB of GPU memory, exceeding the capacity of most consumer-grade GPUs.




Test-Time Adaptation (TTA) \cite{lin2022listen, kim2023sgem,lin2024tta} has gained attention for addressing domain shift problems in ASR models. Traditional unsupervised domain adaptation methods, such as domain adversarial learning \cite{sun2018domain} and data augmentation \cite{hsu2017unsupervised}, rely on access to source data and pre-collected target data. Therefore, these approaches face practical limitations in real-world applications due to storage and computational overhead. TTA methods adapt models in real time using unlabeled samples from the target domain without requiring source data, effectively overcoming these challenges. The framework in \cite{lin2022listen} pioneered single-utterance TTA for ASR and demonstrated strong performance on CTC-based models. The research on \cite{kim2023sgem} builds upon \cite{lin2022listen} by leveraging beam search to explore candidate output logits, further extending the TTA method to autoregressive ASR models and improving performance.

\subsection{Memory Efficient ZO-SGD}

MeZO \cite{malladi2023fine}, along with its low-rank variant LoZO \cite{chen2024enhancing}, are detailed in this section. Unlike mainstream FO optimizers that leverage backpropagation to compute gradients, these approaches belong to the Randomized Vector-wise Gradient Estimation (RGE) class \cite{duchi2015optimal, nesterov2017random, liu2018zeroth}, which operates by estimating gradients through the application of a perturbation $\boldsymbol{z}$ to model parameters:


\begin{equation} \label{eq:1}
\hat{\nabla} \mathcal{L}(\boldsymbol{\theta}) := \frac{\mathcal{L}(\boldsymbol{\theta} + \epsilon \boldsymbol{z}) - \mathcal{L}(\boldsymbol{\theta} - \epsilon \boldsymbol{z})}{2\epsilon} \boldsymbol{z},
\end{equation} 
where $\mathcal{L}$ is the loss function, and $\epsilon$ is the perturbation scale. $\boldsymbol{\theta} = \{ \theta_i \}_{i=1}^I$ represents the set of model weight matrices, with $\theta_i$ denoting the weight matrix of the $i$-th layer. The expectation of this gradient estimate is equal to the expectation of the real gradient, a result derivable from Taylor expansion. The main difference between MeZO and LoZO is the perturbation sampling method to get $\boldsymbol{z}$.

The initial development of MeZO \cite{malladi2023fine} targeted the memory-efficient fine-tuning of LLMs. In contrast to conventional ZO-SGD, MeZO-SGD leverages a random seed trick to sample and regenerate perturbation vectors $z \sim \mathcal{N}(0, 1)$ in Equation (\ref{eq:1}) using the same random seed for random number generator, thereby eliminating the need to store perturbation vectors $\boldsymbol{z}$.  As a result, the memory footprint is halved compared to standard ZO-SGD, facilitating fine-tuning operations within inference memory constraints. The update rule for the parameters is formulated as:

\begin{equation} \label{eq:theta}
\boldsymbol{\theta}^{t+1} := \boldsymbol{\theta}^t - \eta \hat{\nabla} \mathcal{L}(\boldsymbol{\theta}^{t}),
\end{equation} 
for learning rate $\eta$ and model parameter $\boldsymbol{\theta}^t$ at $t$.

As a variant of MeZO, LoZO \cite{chen2024enhancing} distinguishes itself through an alternative perturbation sampling method. This method leverages the inherent low-rank structure in FO gradients during the back-propagation in LLMs \cite{zhao2024galore, hao2024flora}. For any weight matrix $\theta_i \in \mathbb{R}^{m_i \times n_i}$, by sampling two low-rank random matrices \( U_i \in \mathbb{R}^{m_i \times r_i} \) and \( W_i \in \mathbb{R}^{n_i \times r_i} \) with rank $r_i \ll \min\{m_i, n_i\}$ from a normal distribution, it ensures that the random perturbation vector \( z = U_i W_i^\top \) retains a low-rank property.
Integrating LoZO with momentum-based optimization methods necessitates the storage of one random matrix \( U_i \) for each weight matrix. With the random matrix $U_i$ being substantially smaller than $z$, LoZO offers a memory-efficient way to incorporate momentum techniques.

Although MeZO and LoZO have demonstrated efficacy comparable to FO-Adam in fine-tuning LLMs for several downstream tasks, current investigations centered on relatively simple binary and multi-class classification problems (e.g., SST-2, SST-5, CB and, RTE \cite{socher2013recursive, de2019commitmentbank,rte}) with specific prompt templates. These methods face challenges when applied to more intricate generative tasks, such as speech recognition.

\section{Methodology}

In this section, we present ZO-ASR, a novel memory-efficient Zeroth-Order method designed to enhance the performance of the vanilla ZO-SGD algorithm by acquiring more accurate gradient estimates through additional query iterations, as shown in Algorithm \ref{alg:q-SPSA}.

\subsection{Proposed ZO-ASR Method}

ZO-ASR employs a q-query Random Gradient Estimation (q-RGE) \cite{duchi2015optimal} approach, wherein gradient estimates derived from q queries are aggregated for a single gradient update. Conventional MeZO or LoZO optimization strategies typically utilize a single gradient estimate per update. While this approach may yield satisfactory results for simple classification tasks, its efficacy diminishes in more complex speech recognition tasks due to the high variance inherent in such single-estimate updates. Through its q-RGE mechanism, ZO-ASR attains enhanced robustness. The formulation for q-RGE is presented as follows:

\begin{equation} \label{eq:2}
\hat{\nabla} \mathcal{L}^{(q)}(\boldsymbol{\theta}) := \frac{1}{q} \sum_{i=1}^q \frac{\mathcal{L}(\boldsymbol{\theta} + \epsilon \boldsymbol{z}_i) - \mathcal{L}(\boldsymbol{\theta} - \epsilon \boldsymbol{z}_i)}{2\epsilon} \boldsymbol{z}_i.
\end{equation}


ZO-ASR method primarily consists of two stages shown in Algorithm \ref{alg:q-SPSA}. In the initial gradient estimation stage, the algorithm perturbs the parameters designated for fine-tuning and performs a total of $2*q$ forward propagations. It caches $q$ perturbed gradient projections ($\mathtt{proj\_grad}$, i.e., the projection of the true gradient onto a random vector $z$) and the random seeds used for generating these random numbers. In the second model update stage, the algorithm resets the random seeds and utilizes the random number generator to reproduce the perturbed random vectors, which are then used to update the model weights. The model weights are updated using a query step accumulation method, where the gradient estimates from multiple past query steps are accumulated as the final gradient estimate, without being divided by the number of gradient estimations. This intuition is derived from the gradient accumulation method used to simulate large batch sizes.

The perturbation of the model's weight matrix is always performed in-place to conserve GPU memory. If we denote the weights before the loop as $\theta_0$, the model weights in GPU undergo the following sequence: $\theta_0 \to \theta_0 + \epsilon z \to \theta_0 - \epsilon z \to \theta_0.$
This corresponds to a computational process of successively applying the operations: add $\epsilon z$, subtract $2\epsilon z$, and add $\epsilon z$, as shown in Algorithm \ref{alg:q-SPSA}.

\begin{algorithm}[t]
   \caption{ZO-ASR-SGD}
   \label{alg:q-SPSA}
\begin{algorithmic}
   \STATE {\bfseries Input:} parameters $\boldsymbol{\theta} \in \mathbb{R}^d$, loss $\mathcal{L}: \mathbb{R}^d \to \mathbb{R}$, step budget $T$, perturbation scale $\epsilon$, learning rate $\eta$, $q$ query step accumulation
  \FOR{$t = 1, \dots, T$ }
    \item $\mathtt{seeds}, \mathtt{proj\_grads} \leftarrow [ \ ]$ 
    \item \hfill \textit{ \# Stage 1: Gradient Estimation}
    \FOR{$j = 1, \dots, q$ }   
        \item Sample random seed $s$
        \item $\boldsymbol{\theta} \leftarrow \mathtt{Perturb}(\boldsymbol{\theta}, \epsilon, s)$ \hfill \textit{ \# perturb to $\boldsymbol{\theta} + \epsilon \boldsymbol{z}_j$}
        \item $\ell^+ \leftarrow \mathcal{L}(\boldsymbol{\theta})$ \hfill \textit{ \# forward pass and calculate loss}
        \item $\boldsymbol{\theta} \leftarrow \mathtt{Perturb}(\boldsymbol{\theta}, -2\epsilon, s)$ \hfill \textit{ \# perturb to $\boldsymbol{\theta} - \epsilon \boldsymbol{z}_j$}
        \item $\ell^- \leftarrow \mathcal{L} (\boldsymbol{\theta})$ \hfill \textit{ \# forward pass and calculate loss}
        \item $\boldsymbol{\theta} \leftarrow \mathtt{Perturb}(\boldsymbol{\theta}, \epsilon, s)$ \hfill \textit{ \# perturb to $\boldsymbol{\theta}$}
        \item $\mathtt{proj\_grad} \leftarrow (\ell^+ - \ell^-)/(2\epsilon)$ 
        \item $\mathtt{proj\_grads[j]} \leftarrow \mathtt{proj\_grad}$ 
        \item $\mathtt{seeds[j]} \leftarrow s$
    \ENDFOR 
        \item \hfill \textit{ \# Stage 2: Model Update}
      \FOR{$j = 1, \dots, q$ }
        \item Reset seed $\mathtt{seeds[j]}$ \hfill 
        \FOR {$\theta_i \in \boldsymbol{\theta} $} 
          \item $z \sim \mathcal{N}(0, 1)$ \hfill \textit{ \# resample the vector $\boldsymbol{z}_j$}
          \item $\theta_i \leftarrow \theta_i - \eta * \mathtt{proj\_grads[j]} * z$
        \ENDFOR
      \ENDFOR
  \ENDFOR
      \item \hfill \textit{ \# In-place perturb operation}
  \STATE {\bfseries Subroutine} $\mathtt{Perturb}(\boldsymbol{\theta}, \epsilon, s)$
      \item Set random seed s
      \FOR {$\theta_i \in \boldsymbol{\theta} $}
        \item $z \sim \mathcal{N}(0, 1)$
        \item $\theta_i \leftarrow \theta_i + \epsilon z$   
      \ENDFOR
\end{algorithmic}
\end{algorithm}

\subsection{Memory Efficiency}

The memory efficiency of ZO-ASR manifests in two aspects: GPU memory during model training and storage memory for checkpoints.

\textbf{Training GPU Memory.} The primary advantage of the ZO-ASR algorithm is its significant reduction in GPU memory consumption during training. Since only forward propagation is performed throughout the training process, ZO-ASR can achieve fine-tuning with nearly inference-level memory requirements. The maximum additional memory overhead arises from the largest perturbed matrix, which is substantially smaller than the overall model size. 


\textbf{Storage Memory.} ZO-ASR algorithm facilitates memory-efficient checkpoint storage. For instance, when fine-tuning Whisper-Large-v3 on a low-resource language, a checkpoint derived from 100K forward passes can be realized by storing 50K random seeds and their corresponding gradient projections, which implies that storing a checkpoint for a model with 1.55B parameters necessitates less than 1MB of memory. This benefit also extends to the TTA domain. Conventionally, when an ASR model undergoes TTA, it is necessary to retain a version of the source domain model. ZO-ASR obviates this memory expenditure by enabling the model to revert to its source domain state through the storage of only random seeds and gradient projections.

\section{Experiments}

\begin{table*}[htbp]
    \centering
    \renewcommand{\arraystretch}{1.2}
    \caption{Experiments on Whisper-Large-V3 (1550M parameters). WER (\%) is computed across four languages, while CER (\%) is specifically calculated for Thai. In subsequent tables, this formatting convention (\textbf{bold} for best results except FO methods, \textcolor{gray}{gray} for FO-Adam optimizers) will be maintained.}
    \label{tab:whisper}
    \begin{tabular}{l *{5}{c c} c } 
    \toprule
        \multicolumn{1}{l}{\textbf{Language}} & \multicolumn{2}{c}{\textbf{Hindi}} & \multicolumn{2}{c}{\textbf{Thai}} & \multicolumn{2}{c}{\textbf{Galician}} & \multicolumn{2}{c}{\textbf{Lithuanian}} & \multicolumn{2}{c}{\textbf{Portuguese}} & \multirow{2}{*}{\textbf{{Avg. WER}}} \\ 
        \cmidrule(lr){2-3} \cmidrule(lr){4-5} \cmidrule(lr){6-7} \cmidrule(lr){8-9} \cmidrule(lr){10-11}
        \textbf{Method} & \textbf{Loss} & \textbf{WER} & \textbf{Loss} & \textbf{CER} & \textbf{Loss} & \textbf{WER} & \textbf{Loss} & \textbf{WER} & \textbf{Loss} & \textbf{WER} & \\ \midrule
        Zero-Shot & 0.530 & 37.1 & 0.220 & 8.1 & 0.907 & 17.9 & 0.824 & 33.4 & 0.933 & 12.0 & 21.7 \\ \midrule 
        
        MeZO-SGD & 0.245 & 31.2 & \textbf{0.157} & \textbf{6.8} & 0.277 & 17.3 & 0.327 & 31.8 & 0.186 & 12.0 & 19.8 \\ 
        LoZO-SGD & 0.260 & 32.4 & 0.169 & 7.3 & 0.288 & 18.2 & 0.472 & 32.5 & 0.196 & 13.2 & 20.7 \\ 
        ZO-ASR-SGD & \textbf{0.232} & \textbf{30.1} & 0.158 & \textbf{6.8} & \textbf{0.253} & \textbf{16.2} & \textbf{0.321} & \textbf{31.3} & \textbf{0.181} & \textbf{11.6} & \textbf{19.2} \\ \midrule 
        \rowcolor{gray!20}
        FO-Adam (LoRA) & 0.205 & 24.8 & 0.112 & 4.7 & 0.184 & 10.3 & 0.216 & 21.0 & 0.180 & 10.7 & 14.3 \\ 
        \rowcolor{gray!20}
        FO-Adam & 0.200 & 23.4 & 0.110 & 4.7 & 0.228 & 10.1 & 0.198 & 18.4 & 0.177 & 10.4 & 13.4 \\ \bottomrule 
    \end{tabular}
    \vskip -0.1in
\end{table*}


In this section, we evaluate the performance of ZO-ASR against other ZO and FO baselines across different models and tasks.

\textbf{Datasets.} Our experimental datasets encompass two categories of tasks. For the supervised domain adaptation in speech recognition, we selected the Common Voice dataset \cite{ardila2019common} and fine-tuned Whisper-Large-V3 on five low-resource languages: Hindi (HI), Thai (TH), Galician (GL), Lithuanian (LT), and Portuguese (PT). Due to the limited data, the training and validation sets were merged, and evaluation was performed exclusively on the test sets. For the unsupervised TTA, we followed the experimental setup of \cite{lin2022listen} and utilized the test sets from LibriSpeech \cite{panayotov2015librispeech} test-other, TED-LIUM v3 \cite{hernandez2018ted}, and CHiME3 \cite{barker2015third} dataset. To assess the robustness of TTA using ZO-ASR in noisy environments, we introduced varying magnitudes of Gaussian noise ( $\sigma = 0, 5e-3, 1e-2$) to the LibriSpeech test-other dataset. 

\textbf{Models.} Our experiments investigated the performance of ZO-ASR across different model architectures and sizes. This included the Whisper series (Tiny, Base, and Large-v3) based on an encoder-decoder Transformer architecture, and Wav2Vec2-Base based on an encoder-only architecture.

\textbf{Implementation Details.} For a fair comparison, the experiments involving ZO optimization are conducted by
controlling for an equivalent number of total forward passes. All experiments were conducted on NVIDIA 3090 and A800 GPUs. Diverging from existing research on LLMs, ASR models do not support instruction prompts. Consequently, all experiments were performed in a prompt-free setting, directly replacing the original BP-based optimizer (e.g., FO-Adam or FO-SGD) with ZO-ASR. Considering the potential impact of small batch size on ZO optimization \cite{gautam2024variance}, all comparative experiments involving ZO-based methods utilized an identical and large enough batch size of 24.

\textbf{Metrics.} The evaluation of our experiments focused on accuracy and computational cost. For ASR models, accuracy was quantified by the Word Error Rate (WER) or Character Error Rate (CER) in speech recognition, where lower WER or CER values signify higher accuracy. Computational cost was assessed in terms of the number of training steps and latency.

\subsection{Results on Supervised Domain Adaptation}

\textbf{The performance of ZO-ASR consistently surpasses zero-shot and other ZO baselines.}
As demonstrated in Table \ref{tab:whisper}, ZO-ASR achieves lower WER and test losses across the majority of tasks compared to alternative ZO optimization methods. In comparison to zero-shot settings, ZO-ASR achieves up to 18.9\% reduction for Hindi language and 11.5\% average reduction in WER utilizing only forward propagation and inference-level memory. Nevertheless, its performance remains considerably inferior to FO-Adam. This discrepancy arises because the convergence of ZO optimization is contingent upon the number of model parameters. 1.55 billion parameters of Whisper-Large-v3 attenuate the convergence efficacy of ZO-ASR.




\setlength{\cmidrulewidth}{0.1mm} 
\renewcommand{\arraystretch}{1.0} 
\begin{table}[tbp]
    \caption{Comparison of pre-training and fine-tuning hours (h) for Whisper-Large-V3 across different languages, including the Ratio of pre-training to fine-tuning Time.}
    \label{tab:time_comparison}
    \centering
    \begin{tabular}{lccc} 
        \toprule 
        \textbf{Language} & \textbf{Pre-Training} & \textbf{Fine-Tuning} & \textbf{Ratio} \\ \midrule
        Hindi & 12h & 8h & 1.5 \\
        Lithuanian & 67h & 11h & 6.1 \\
        Thai & 226h & 45h & 5.0 \\
        Portuguese & 8573h & 30h & 285.8 \\
        \bottomrule 
    \end{tabular}
\end{table}

\textbf{The volume of data influences the performance of ZO fine-tuning for domain adaptation in low-resource languages.}
ZO-based fine-tuning methods exhibit varied performance across different low-resource languages, primarily due to the influence of the data used to train the pre-trained model. Table \ref{tab:time_comparison} illustrates the data volumes utilized by the Whisper model during its pre-training phase and our subsequent fine-tuning phase. When the volume of fine-tuning data closely approximates that of the pre-training data (e.g., for Hindi), ZO fine-tuning yields substantial improvements, achieving a word error rate reduction of nearly 18.9\% relative to the zero-shot baseline. Conversely, when the fine-tuning data volume is considerably smaller than the pre-training data volume (e.g., for Portuguese), only the ZO-ASR method effectively facilitates domain adaptation. In contrast, MeZO and LoZO result in performance degradation, thereby highlighting the superior robustness of ZO-ASR.

\setlength{\cmidrulewidth}{0.1mm} 
\renewcommand{\arraystretch}{1.0} 
\begin{table}[tbp]
    \centering
    \caption{Comparison of WER (\%) across different Whisper model sizes and fine-tuning methods on Portuguese language. The tiny subscript number indicates the Relative WER Reduction (\%).}
    \label{tab:diff_size}
    \begin{tabular}{llll}
        \toprule
        \textbf{Method} & \textbf{Large-V3} & \textbf{Base} & \textbf{Tiny} \\
        & \textbf{(1550MB)} & \textbf{(74MB)} & \textbf{(39MB)} \\
        \midrule
        Zero-Shot       & 12.0     & 36.8     & 53.2     \\
        MeZO-SGD        & \werSub{12.0}{ 0}     & \werSub{31.7}{ -13.9}  & \werSub{40.9}{ -23.1}  \\
        ZO-ASR-SGD      & \werSub{11.6}{ -3.3}   & \werSub{30.1}{ -18.2}  & \werSub{39.8}{ -25.2}  \\
        FO-Adam    & \werSub{10.4}{ -13.3}  & \werSub{22.1}{ -39.9}  & \werSub{31.3}{ -41.2}  \\
        \bottomrule
    \end{tabular}
\end{table}

\textbf{The effectiveness of ZO-ASR is more pronounced when fine-tuning smaller models.} As shown in Table \ref{tab:diff_size}, as the model size decreases, ZO-ASR-SGD achieves a Word Error Rate (WER) reduction of 0.4\%, 6.7\%, and 13.4\% compared to the zero-shot baseline. This corresponds to a Relative WER Reduction of 3.3\%, 18.2\%, and 25.2\%, respectively. These results indicate that for a given model architecture, ZO-ASR yields a more significant performance improvement when fine-tuning models with a smaller number of parameters. Nevertheless, for both the Whisper-Tiny and Whisper-Base models, a noticeable performance gap persists between ZO-ASR and First-Order (FO) Adam.

\subsection{Results on Unsupervised Test-time Adaptation}

\textbf{ZO-based methods achieve performance with a moderate degradation compared to FO-Adam, primarily attributable to a substantial reduction in the number of trainable parameters, as shown in table \ref{tab:tta}.} Firstly, Wav2Vec2 possesses a parameter count of only 95 MB, considerably smaller than the 1.55 billion parameters of Whisper-Large-v3. Secondly, the TTA methodology proposed by \cite{lin2022listen} exclusively fine-tunes the CNN layers of the feature extractor and the LayerNorm layers. This further reduces the number of tunable parameters to a count that is smaller than the size of the Whisper-Tiny or Whisper-Base. Besides, LoZO's low-rank matrix decomposition technique is designed to perform low-rank sampling exclusively on two-dimensional matrices, while both the CNN and LayerNorm layers within ASR models can be considered as one-dimensional (1D) matrices. Consequently, when processing 1D matrices, LoZO defaults to the same Gaussian sampling mechanism employed by MeZO. As a result, LoZO and MeZO exhibit identical performance in TTA experiment, and LoZO's results are omitted from Table \ref{tab:tta}. Furthermore, ZO-based approaches can store model parameter updates using random seeds and projected gradients. This technique virtually eliminates the storage overhead typically required for backing up the original model parameters in TTA tasks.

\begin{table}[tbp]
    \centering
    \renewcommand{\arraystretch}{1.2}
    \caption{WER (\%) of Wav2Vec2-Base (95M parameters) adapted to LibriSpeech (LS), TEDLIUM (TED), and CHiME3 datasets. LS is added with varying noise levels ($\sigma = 0, 5e-3, 1e-2$). We run ZO for 640 total forward passes and FT for 10 training steps.}
    \label{tab:tta}
    \begin{tabular}{p{1.90cm} p{0.70cm} p{0.75cm} p{0.75cm} >{\centering\arraybackslash}p{0.90cm} >{\centering\arraybackslash}p{1.2cm} }
    \toprule
        \multicolumn{1}{c}{} & \multicolumn{3}{c}{\textbf{LS ($\sigma$)}} & \multirow{2}{*}{\textbf{TED}} & \multirow{2}{*}{\textbf{CHiME3}}\\
        \cmidrule(lr){2-4}
        \textbf{ Noise $\sigma$} & 0 & 5e{-3} & 1e{-2} \\ \midrule
        
        Zero-Shot & 8.6 & 13.9 & 24.5 & 13.2 & 31.2 \\ \midrule
        MeZO-SGD & 7.7 & 11.3 & 17.8 & 12.6 & 26.3 \\
        ZO-ASR-SGD & 7.7 & 11.3 & 17.8 & 12.5 & 26.3 \\
        \midrule
        \rowcolor{gray!20}
        FO-Adam & 7.3 & 10.8 & 16.5 & 11.9 & 25.0 \\
        \bottomrule
    \end{tabular}
\end{table}

\begin{table}[tbp]
    \centering
    \renewcommand{\arraystretch}{1.0}
    \caption{Performance of Wav2Vec2-Base on Librispeech test-other dataset with varying magnitudes of Gaussian noise ($\sigma = 0, 5e-3, 1e-2$) during adaptation. WER (\%) and TTA time (seconds) for a 1-second utterance are computed at each TTA step. FO-Adam adopts 10-steps TTA.}
    \label{tab:tta-100steps}
    \begin{tabular}{l c c c c }
    \toprule

        \multicolumn{1}{c}{} & \multicolumn{3}{c}{\textbf{WER}} & \multirow{2}{*}{\textbf{Adaptation time}}\\
        \cmidrule(lr){2-4}

        \textbf{ Noise $\sigma$} & 0 & 5e{-3} & 1e{-2} \\ \midrule
        
        Zero-Shot & 8.6 & 13.9 & 24.5 & -- \\
        \midrule
        1-step  & 8.5 & 13.7 & 23.7 & 0.06s \\
        3-steps & 8.4 & 13.2 & 22.3 & 0.18s \\
        5-steps & 8.3 & 12.8 & 21.5 & 0.31s \\
        10-steps & 8.1 & 12.3 & 20.1 & 0.61s \\
        30-steps & 7.8 & 11.5 & 18.1 & 1.84s \\
        50-steps & 7.7 & 11.3 & 17.5 & 3.07s \\
        100-steps & \textbf{7.5} & \textbf{11.1} & \textbf{17.1} & 6.14s \\
        \midrule
        \rowcolor{gray!20}
        FO-Adam & 7.3 & 10.6 & 16.1 & 0.12s \\
    \bottomrule
    \end{tabular}
\end{table}

\textbf{The primary bottleneck associated with ZO fine-tuning stems from its significantly higher latency compared to FO-Adam.} As illustrated in Table \ref{tab:tta-100steps}, for ZO-ASR to attain performance approaching those of FO methods, it necessitates a time overhead exceeding 50-fold. Despite this limitation, ZO-ASR, being a BP-free approach, holds potential for deployment on edge devices where gradients may be inaccessible, or for application to quantized models.



        

\subsection{Ablation Study}
We conducted an ablation study on query numbers to investigate whether directly increasing the number of queries to $q$ could improve the performance of ZO-ASR. The experimental results are shown in Table \ref{tab:abla}. Scaling the query numbers directly achieves good performance when $q=8$, but as $q$ increases further, the performance deteriorates due to excessively high learning rates.

\begin{table}[tbp]
    \centering
    \renewcommand{\arraystretch}{1.2}
    \caption{Experiments on Whisper-Large-V3 on Hindi and Lithuanian languages with different query numbers.}
    \label{tab:abla}
    \begin{tabular}{c c c c c}
    \toprule
        \multicolumn{1}{l}{} & \multicolumn{2}{c}{\textbf{Hindi}} & \multicolumn{2}{c}{\textbf{Lithuanian}} \\
        \cmidrule(lr){2-3} \cmidrule(lr){4-5} 
        \textbf{Query Number} & \textbf{Loss} & \textbf{WER} & \textbf{Loss} & \textbf{WER} \\ \midrule
        Zero-Shot & 0.530 & 37.1 & 0.824 & 33.4 \\
        \midrule
        1 & 0.245 & 31.2 & 0.330 & 32.2 \\ 
        2 & 0.241 & 30.7 & 0.326 & 31.5 \\ 
        4 & 0.238 & 30.6 & 0.326 & 31.5 \\ 
        8 & \textbf{0.232} & \textbf{30.1} & \textbf{0.321} & \textbf{31.3} \\ 
        16 & 0.236 & 30.3 & 0.328 & 31.7 \\  \midrule
        \rowcolor{gray!20}
        FO-Adam & 0.200 & 23.4 & 0.189 & 18.4 \\
        \bottomrule
    \end{tabular}
\end{table}

\subsection{Memory Efficiency}

Table \ref{tab:memory} illustrates the GPU peak memory requirements for ZO-ASR and BP-based optimizers when performing full-parameter or LoRA fine-tuning of Whisper-Large-V3. ZO-ASR achieves fine-tuning with memory consumption approaching inference in both configurations, requiring less than 25.5\% of the memory utilized by the BP-based optimizer. Conversely, the BP-based optimizer, whether employed for full fine-tuning or with the LoRA method, necessitates GPU memory several times that of inference, exceeding the memory capacity of a single consumer-grade GPU.


\setlength{\cmidrulewidth}{0.1mm} 
\renewcommand{\arraystretch}{1.1} 
\begin{table}[t]
    \caption{GPU Memory consumption of FO and ZO-ASR methods to finetune Whisper-Large-V3 on Thai language, with FP32 precision and batch size of 4.}
    \label{tab:memory}
    \centering
    \begin{tabular}{lcc} 
        \toprule 
        \textbf{Optimizer} & \textbf{w/ LoRA} & \textbf{w/o LoRA} \\ \midrule
        FO-SGD & 37.7 GB & 43.9 GB \\
        FO-Adam & 37.8 GB & 56.4 GB \\
        ZO-ASR-SGD & \textbf{9.5 GB} & \textbf{9.6 GB} \\
        \midrule
        Inference & \multicolumn{2}{c}{9.5GB} \\ 
        \bottomrule 
    \end{tabular}
\end{table}

\section{More Discussions}

\subsection{Improvement Strategies}

Several studies have explored methods to enhance ZO optimization performance, such as hybrid FO-ZO training \cite{gautam2024variance}, momentum-based optimizers \cite{chen2019zo}, and the use of second-order Hessian information \cite{zhao2024second}. These approaches have proven effective on various tasks involving LLMs and suggest promising avenues for applying ZO optimization to ASR. However, this improved performance comes at a cost: these methods typically require more GPU memory than baselines like MeZO or LoZO. For example, hybrid training needs extra storage for activations, while momentum and second-order methods require additional memory for their corresponding matrices.

\subsection{Future Optimization}

A promising future direction is to accelerate the ZO optimization process by focusing on its inference stage. The latency of ZO primarily stems from the computational overhead of the forward pass; thus, speeding up this step would significantly reduce overall latency. Existing work on LLMs inference acceleration has already achieved remarkable success. Techniques like vLLM \cite{kwon2023efficient} and SGLang \cite{zheng2024sglang} can boost LLMs inference speed by over tenfold compared to the baseline \texttt{huggingface.transformers} implementation. Therefore, adapting these techniques to ZO optimization—for instance, by enabling partial parameter perturbation post-deployment—could substantially reduce the process's latency.

\section{Conclusion}

This paper introduces ZO-ASR, a novel BP-free and memory-efficient method for fine-tuning ASR foundation models. ZO-ASR demonstrates the capability to reduce WER by up to 18.9\% for Whisper-Large-v3 in low-resource language recognition tasks, and it achieves performance with a moderate degradation compared to FO-Adam in unsupervised TTA tasks on Wav2Vec2-Base. Eliminating back-propagation and enabling fine-tuning with inference-level memory, ZO-ASR extends the applicability of ASR model fine-tuning to resource-constrained settings and offers a potential solution for the fine-tuning of models post-quantization or after deployment on edge devices. Current limitations primarily pertain to convergence speed and overall performance. Particularly when fine-tuning models with a substantial number of parameters, a significant performance gap persists between ZO-ASR and FO-Adam.

\section*{Acknowledgment}

This work was supported by the Science and Technology Innovation (STI) 2030-Major Project (2022ZD0208700), National Natural Science Foundation of China  (No. 62206171 and No. U23B2018), Shanghai Municipal Science and Technology Major Project under Grant 2021SHZDZX0102 and Yangtze River Delta Science and Technology Innovation Community Joint Research Project (2024CSJGG01100).

\bibliographystyle{IEEEtran}
\bibliography{main}

\end{document}